\makeatletter
\def\input@path{{lib/}}
\makeatother
\documentclass[acmsmall,screen,nonacm,natbib=false]{acmart}

\usepackage[utf8]{inputenc}
\usepackage[T1]{fontenc}

\usepackage[
  backend=biber,
  style=acmnumeric,
  sorting=nty,
  maxbibnames=99,
  giveninits=true,
  bibencoding=utf8
]{biblatex}
\AtEveryBibitem{\clearfield{keywords}}

\usepackage{url}
\usepackage{xurl}
\usepackage{etoolbox}
\usepackage{siunitx}
\usepackage{wrapfig}
\usepackage{multirow}
\usepackage{makecell}
\usepackage{graphicx}
\usepackage{subcaption}
\usepackage{textcomp}
\usepackage[table]{xcolor}
\usepackage{multirow}
\usepackage{tabularx} 
\usepackage{soul} 
\usepackage{orcidlink}
\usepackage{listings}
\usepackage{pythonhighlight}
\usepackage{algorithm}
\usepackage{algpseudocode}
\usepackage{amsmath}
\usepackage{paralist}

\providecommand{\xurl}[1]{\url{#1}}
\AtBeginDocument{%
  }

\renewcommand\footnotetextcopyrightpermission[1]{}

\newcommand{\up}{\vspace*{-1em}}
\newcommand{\upp}{\vspace*{-0.5em}}

\begin{document}

\title{Hybrid Quantum-HPC Middleware Systems for Adaptive Resource, Workload and Task Management}

\author{Pradeep Mantha}
\orcid{0000-0003-2664-7737}
\affiliation{%
  \institution{Ludwig Maximilian University}
  \city{Munich}
  \country{Germany}
}
\email{pradeep.mantha@campus.lmu.de}

\author{Florian J. Kiwit}
\orcid{0009-0000-4065-1535}
\affiliation{%
  \institution{BMW Group}
  \city{Munich}
  \country{Germany}
}
\affiliation{%
  \institution{Ludwig Maximilian University}
  \city{Munich}
  \country{Germany}
}
\email{f.kiwit@campus.lmu.de}

\author{Nishant Saurabh}
\orcid{0000-0002-1926-4693}
\affiliation{%
  \institution{Department of Information and Computing Sciences, Utrecht University}
  \country{The Netherlands}
}
\email{n.saurabh@uu.nl}

\author{Shantenu Jha}
\orcid{0000-0002-5040-026X}
\affiliation{%
  \institution{Rutgers University-New Brunswick}
  \state{New Jersey}
  \country{USA}
}
\affiliation{%
  \institution{Princeton Plasma Physics Laboratory}
  \state{New Jersey}
  \country{USA}
}
\affiliation{%
  \institution{Princeton University}
  \state{New Jersey}
  \country{USA}
}
\email{shantenu.jha@rutgers.edu}

\author{Andre Luckow}
\orcid{0000-0002-1225-4062}
\affiliation{%
  \institution{BMW Group}
  \city{Munich}
  \country{Germany}
}
\affiliation{%
  \institution{Ludwig Maximilian University}
  \city{Munich}
  \country{Germany}
}
\email{andre.luckow@bmwgroup.com}

\renewcommand{\shortauthors}{Mantha, et al.}

\begin{abstract}
Hybrid quantum-classical applications pose significant resource management challenges due to heterogeneity and dynamism in both infrastructure and workloads. Quantum-HPC environments integrate scarce and variable quantum processing units (QPUs) with diverse classical resources (CPUs and GPUs), each exhibiting distinct performance characteristics and usage constraints. At the same time, applications span a spectrum of coupling patterns--from tightly coupled execution to loosely coupled task parallelism---and exhibit stage-dependent, adaptive resource requirements across complex workflows. Traditional HPC schedulers, which operate at coarse job-level granularity with early binding, lack visibility into application semantics and are unable to respond to fluctuating resource availability or evolving workload demands at runtime. This paper presents a middleware-based approach for adaptive resource, workload, and task management in hybrid quantum-HPC systems. We make four primary contributions. First, we introduce a conceptual four-layer middleware architecture that decomposes management across workflow, workload, task, and resource levels, enabling application-aware scheduling over heterogeneous quantum-HPC resources. Second, we define a set of execution motifs that capture the interaction and coupling characteristics of hybrid quantum-classical applications; these motifs are realized as quantum mini-apps to support systematic workload characterization. Third, we present Pilot-Quantum, a middleware framework built on the pilot abstraction that enables late binding and dynamic resource allocation, allowing application-level scheduling to adapt to resource and workload dynamics at runtime. Fourth, we introduce Q-Dreamer, a performance modeling toolkit that provides reusable components for informed workload partitioning and resource allocation decisions, including a circuit-cutting resource optimizer that analytically derives optimal partitioning strategies. Evaluation on heterogeneous HPC platforms (Perlmutter, NVIDIA DGX with H100/B200 GPUs) demonstrates that Pilot-Quantum achieves efficient multi-backend orchestration across CPUs, GPUs, and QPUs for diverse execution motifs, including parallel circuit execution, circuit cutting, and multi-stage QML workflows. Q-Dreamer accurately predicts optimal circuit cutting configurations with up to $82\%$ accuracy.

\end{abstract}



\maketitle

\section{Introduction} \label{sec:introduction}

Quantum computing continues to advance rapidly, with recent milestones in hardware~\cite{ransfordHelios2025, googleErgodicity2025, microsoftParity2025} and error correction~\cite{acharya2024quantumerrorcorrectionsurface, rodriguezMagic2025} supporting the transition from noisy intermediate-scale quantum computing (NISQ) toward fault tolerant quantum computing (FTQC)~\cite{eisert2025mindgapsfraughtroad}. These developments are expected to broaden scientific applications and later enable industrial value through quantum simulation, machine learning, and optimization~\cite{bayerstadler2021industry,RIOFRIO2026181}. 

Realizing QC's potential requires integration with classical high-performance computing (HPC)~\cite{ALEXEEV2024666,BECK202411, mohseni2025buildquantumsupercomputerscaling, hpcqc,11071588}. By integration, we do not mean that one paradigm subsumes or merely accelerates the other; rather, quantum and classical systems each contribute complementary capabilities to a shared problem. QPUs address computations where quantum computing offers an advantage (e.\,g., simulation), while HPC systems provide the large-scale data processing, numerical optimization, and error mitigation that quantum algorithms require. Current hardware limitations reinforce this complementarity: limited qubit counts, high error rates, and restricted gate sets necessitate hybrid decompositions in which both quantum and classical components are essential. This mutual dependence will intensify in the FTQC era, where quantum error correction itself demands substantial classical co-processing. Quantum-HPC integration thus spans a spectrum of coupling modes, from tightly-coupled co-processing (e.\,g., error correction) to loosely-coupled task-parallel workflows (e.\,g., variational algorithms, circuit cutting)~\cite{ALEXEEV2024666, saurabh2023conceptual}.

Quantum-HPC systems introduce resource management challenges driven by increased heterogeneity and dynamism in resources and applications. (i)~Resource heterogeneity: quantum-HPC environments must integrate QPUs of different modalities (superconducting, trapped-ion, neutral-atom) with diverse classical accelerators (GPUs, TPUs, FPGAs), each with distinct performance characteristics and programming models; moreover, classical resources are plentiful and standardized while QPUs remain scarce, expensive, and subject to performance variability from calibration cycles and environmental fluctuations. (ii)~Application heterogeneity: quantum-HPC applications exhibit varying coupling requirements between quantum and classical components, must navigate quality-performance tradeoffs where quantum computation can be substituted with classical alternatives (e.\,g., circuit cutting), and demand dynamic resource allocation across workflow stages.

Quantum-HPC applications exhibit diverse resource management requirements: tightly-coupled scenarios (e.\,g., error correction, dynamic circuits) require co-allocation of QPU and classical resources with fixed assignments for microsecond-latency feedback, whereas loosely-coupled scenarios (e.\,g., variational algorithms, circuit cutting) benefit from late binding to enable load balancing and adaptation to shifting resource availability. Neither quantum programming frameworks (e.\,g., Qiskit and Pennylane), which focus on single-backend circuit execution~\cite{11071588,shehata2025bridgingparadigmsdesigninghpcquantum}, nor traditional HPC schedulers, which enforce early binding and exclusive resource access at coarse job granularity~\cite{10.1145/3177851}, address these requirements. This exposes a significant gap in middleware that abstracts heterogeneous quantum-HPC resource pools and enables adaptive, application-level scheduling across them.

Given these challenges, how can middleware enable application-level scheduling for hybrid quantum-HPC workloads? We focus on three dimensions:
\emph{(i)~Characterization:} What are the recurring execution patterns in quantum-HPC applications that middleware must support, and how do these patterns inform scheduling and resource allocation strategies?
\emph{(ii)~Abstractions and Middleware:} What abstractions and middleware systems enable flexible, application-level scheduling across heterogeneous quantum-HPC infrastructure?
\emph{(iii)~Scheduling:} How can applications make informed runtime decisions about resource allocation, workload partitioning and placement?

We address these challenges through four contributions: (i)~a conceptual four-layer middleware architecture (section~\ref{sec:state-of-art}) that decomposes resource management into workflow (L4), workload (L3), task (L2), and resource (L1) levels~\cite{saurabh2023conceptual}; (ii)~execution motifs and quantum mini-apps~\cite{saurabh2024quantum} that formalize interaction and coupling characteristics for systematic workload characterization (section~\ref{sec:motifs}); (iii)~Pilot-Quantum~\cite{mantha2024pilot, 10.1145/3177851}, a middleware built on the pilot abstraction that enables adaptive, application-level scheduling by integrating application semantics with dynamic system state (section~\ref{sec:pilot-quantum}); and (iv)~Q-Dreamer, a workload partitioning and execution optimization toolkit whose Circuit Cutting Resource Optimizer uses calibrated analytical models to determine optimal partitioning strategies (section~\ref{sec:q-dreamer}). Evaluation on heterogeneous HPC platforms (Perlmutter, NVIDIA DGX with H100/B200 GPUs) shows that Pilot-Quantum efficiently orchestrates tasks across CPUs, GPUs, and QPUs, while Q-Dreamer accurately predicts optimal circuit cutting configurations with up to $82\%$ accuracy in cut selection (section~\ref{sec:eval}).

\section{Quantum-HPC Integration Modes}
\label{sec:background}

\begin{figure}
  \centering
  \includegraphics[width=0.8\linewidth]{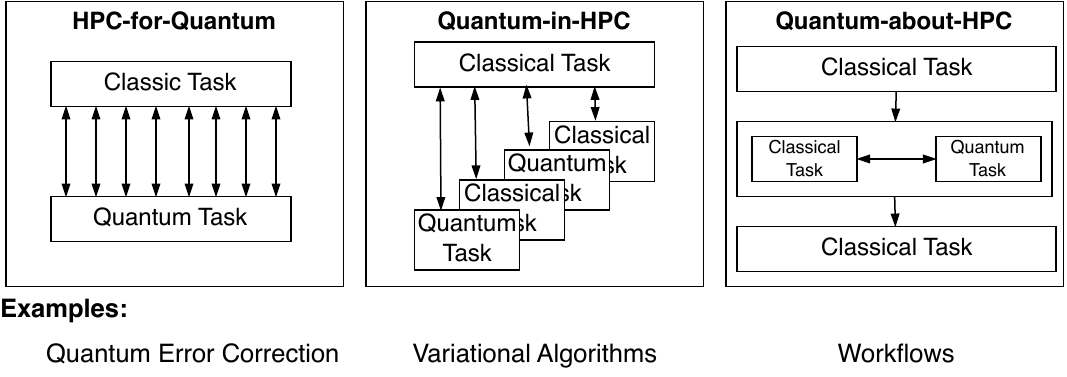}
  \caption{\textbf{Quantum-HPC Integration Patterns:} HPC-for-Quantum requires interactions within the coherence time of the QPU, Quantum-in-HPC a mix of classical and quantum tasks that need to be orchestrated, Quantum-about-HPC connects composable tasks to workflows.}
  \label{fig:quantum-hpc-integration-patterns}  
\end{figure}

Integration modes are categorized as \emph{Quantum-x-HPC} or \emph{HPC-x-Quantum} based on the strength of coupling between quantum and classical tasks~\cite{ALEXEEV2024666, saurabh2023conceptual}. We identified three integration modes~\cite{saurabh2023conceptual} (see figure~\ref{fig:quantum-hpc-integration-patterns}): \emph{HPC-for-Quantum}, with strong coupling optimizing quantum-classical interactions (e.\,g., error mitigation~\cite{ella2023quantumclassicalprocessingbenchmarkingpulselevel}, error correction~\cite{hong2024entanglinglogicalqubitsbreakeven, Bluvstein_2023}, dynamic circuits~\cite{PhysRevLett.127.100501}); \emph{Quantum-in-HPC} emphasizes loosely-coupled parallelism between classical and quantum components found in hybrid applications, such as VQA~\cite{vqa}, to accelerate HPC applications; and \emph{Quantum-about-HPC}, with loose coupling of quantum tasks into HPC applications and workflows~\cite{https://doi.org/10.48550/arxiv.2101.06250}.

These integration modes can be realized through different QPU access modes~\cite{elsharkawy2023integration}: (i)~dedicated node-attached QPUs co-located with classical computation, enabling tight coupling required for HPC-for-Quantum scenarios; (ii)~shared remote QPUs accessed through cloud APIs with multi-tenant queuing; and (iii)~session-based access, where cloud providers offer temporary dedicated QPU access with co-located classical execution (e.\,g., Qiskit Runtime Sessions, Amazon Braket Jobs). While HPC-for-Quantum requires dedicated access (i), Quantum-in-HPC and Quantum-about-HPC can leverage all three access modes depending on coupling requirements.

\emph{HPC-for-Quantum} deploys classical HPC resources for computationally intensive, tightly-coupled support of calibration, real-time control, error correction, and circuit execution within qubit coherence times. This includes pulse-level control with feedback loops faster than decoherence times~\cite{ella2023quantumclassicalprocessingbenchmarkingpulselevel}, e.\,g., error correction with real-time syndrome decoding~\cite{mohseni2025buildquantumsupercomputerscaling}, dynamic circuits with mid-circuit measurements~\cite{PhysRevLett.127.100501}, and measurement-based quantum computation. These capabilities provide essential building blocks for Quantum-in-HPC and Quantum-about-HPC modes.

\begin{table*}[t]
  \centering
  \caption{{\bf Quantum-HPC Application Scenarios:} {\normalfont Application scenarios organized by integration mode: HPC-for-Quantum (classical HPC enables quantum tasks), Quantum-in-HPC (quantum processing embedded in HPC applications), and Quantum-about-HPC (quantum components in end-to-end workflows).}}
  \tiny
  \setlength{\tabcolsep}{1pt}
  \renewcommand{\arraystretch}{1.3}
    
  \begin{tabular}{|>{\raggedright\arraybackslash}p{2.95cm}|>{\raggedright\arraybackslash}p{3.2cm}|>{\raggedright\arraybackslash}p{0.9cm}|>{\raggedright\arraybackslash}p{1.cm}|>{\raggedright\arraybackslash}p{1.5cm}|>{\raggedright\arraybackslash}p{2.1cm}|>{\raggedright\arraybackslash}p{1.7cm}|}
  \hline
  \textbf{Scenario} & \textbf{Description} & \textbf{Coup\-ling} & \textbf{Category} & \textbf{Application structure} & \textbf{Classical Task} & \textbf{Quantum Task}\\
  \hline
  \multicolumn{7}{|c|}{\cellcolor{gray!20}\textbf{HPC-for-Quantum}}\\
  \hline
  Calibration~\cite{ella2023quantumclassicalprocessingbenchmarkingpulselevel} & Real-time calibration and parameter optimization & Tight & Near-term/FTQC & Accelerators & Feedback loops & Calibration circuits (e.\,g., Rabi)\\
  \hline
  Quantum control~\cite{ella2023quantumclassicalprocessingbenchmarkingpulselevel,niu2019universal} & Pulse optimization, gate tuning, adiabatic/diabatic scheduling & Tight & Near-term/FTQC & Accelerators & Bayesian optimizer, RL & Circuit execution, annealing\\
  \hline
  Error Correction~\cite{PhysRevLett.108.180501} & Error-correcting codes (surface, LDPC) for detection and correction & Tight & FTQC & Accelerators & Syndrome decoding & Circuit execution\\
  \hline
  Error Mitigation~\cite{PhysRevLett.119.180509} & Noise reduction via ZNE~\cite{PhysRevLett.119.180509}, PEC~\cite{vandenberg2023pec}, learning~\cite{Czarnik2021errormitigation} & Medium & Near-term & Accelerators & Circuit simulation, extrapolation & Circuit execution\\
  \hline
  Dynamic circuits~\cite{PhysRevLett.127.100501} & Mid-circuit measurements enabling real-time conditional operations & Tight & Near-term/FTQC & Accelerators & Real-time feedback & Conditional gates, resets\\
  \hline
  Circuit cutting~\cite{PhysRevLett.125.150504} & Decomposing large circuits into smaller subcircuits for distribution & Medium & Near-term & Task Parallelism & Decomposition, reconstruction & Subcircuit execution\\
  \hline
  Classical simulation~\cite{cuquantum} & HPC methods to simulate quantum computers & -- & Classical & Task Parallelism, Accelerators & Statevector, tensor network, density matrix & n/a\\
  \hline
  \multicolumn{7}{|c|}{\cellcolor{gray!20}\textbf{Quantum-in-HPC}}\\
  \hline
  Hamiltonian Simulation~\cite{doi:10.1126/science.273.5278.1073} & Time evolution of Schr\"odinger's equation & -- & FTQC & -- & Pre-/post-processing & Hamiltonian simulation\\
  \hline
  Quantum Phase Estimation (QPE)~\cite{qpe} & Extract eigenvalues and eigenstates from a Hamiltonian & Medium & FTQC & Task Parallelism & Inverse QFT, classical analysis & QPE circuit\\
  \hline
  Variational Quantum Algorithms (VQE, QAOA, QNN)~\cite{vqa} & Parameterized quantum circuits with iterative classical optimization & Medium & Near-term & Task Parallelism & Optimization, gradient estimation & Circuit expectation value\\
  \hline
  Subspace Quantum Diagonalization (SQD)~\cite{kanno2023quantumselectedconfigurationinteractionclassical, shirakawa2025closedloopcalculationselectronicstructure} & Hybrid workflow: PQC sampling, subspace selection, diagonalization & Medium & Near-term & Task Parallelism & Subspace construction, diagonalization & Circuit sampling\\
  \hline
  Variational Imaginary Time Evolution (VITE)~\cite{qite} & Ground state via imaginary-time evolution with parameterized ansatz & Medium & Near-term & Task Parallelism & Linear system solving & Circuit expectation value\\
  \hline
  Quantum Machine Learning (QML)~\cite{qcbm_quantum} & QCBM, QGANs with classical or quantum discriminator & Medium & Near-term & Task Parallelism, Accelerators & Optimizer, discriminator & Circuit sampling, expectation value\\
  \hline
  Decoded Quantum Interferometry (DQI)~\cite{dqi-nature2025,sabater2025solvingindustrialintegerlinear} & QFT-based interference mapping structured instances to decoding & Medium & FTQC & Task Parallelism & Problem mapping, decoder (e.\,g., LDPC) & Interferometry circuit\\
  \hline
  \multicolumn{7}{|c|}{\cellcolor{gray!20}\textbf{Quantum-about-HPC}}\\
  \hline
  Classical preprocessing & Encode classical data into quantum state~\cite{Schuld2021} & Loose & Near-term/FTQC & Workflow & Data embedding, encoding & Application-dependent\\
  \hline
  Classical post-processing & Extract and process QPU measurement results & Loose & Near-term/FTQC & Workflow & Expectation value processing & Application-dependent\\
  \hline
  Hyperparameter opt.~\cite{9973678} & Select optimal quantum kernel parameters & Loose & Near-term & Ensemble & Parameter selection & Application-dependent\\
  \hline
  AI workflows & QML for simulation input and property prediction~\cite{generative_molecule_design_2022,https://doi.org/10.48550/arxiv.2101.06250} & Loose & Near-term & Workflow & Simulation loop, optimization & Application-dependent\\
  \hline
  Warm starting~\cite{Egger_2021} & Initialize quantum algorithm with classical solution & Loose & Near-term & Workflow & Heuristics (MILP, CPLEX) & Application-dependent\\
  \hline
  Hybrid QMC~\cite{google_qmc} & QPU supplies overlap estimates to unbias constrained QMC; classical performs time evolution and sampling & Medium & Near-term & Task Parallelism & Sample generation, time evolution & Overlap estimation\\
  \hline
  Quantum-Quantum coupling~\cite{Vazquez_ibm_circuit_cutting_2qpus,distributed_qc_optical2025} & Distributed computation across networked QPUs via entanglement or classical links & Tight & Near-term/FTQC & Task Parallelism & Entanglement distribution, circuit cutting, reconstruction & Distributed circuits, Hamiltonian simulation\\
  \hline
  \end{tabular}
  \label{tab:quantum-hpc-scenarios}
\end{table*}

The \emph{Quantum-in-HPC} mode embeds quantum processing capabilities within HPC applications (table~\ref{tab:quantum-hpc-scenarios}). Variational quantum algorithms (VQAs) represent a prime example of this integration mode~\cite{vqa}. VQAs encompass a broad family of algorithms, including the variational quantum eigensolver (VQE), the quantum approximate optimization algorithm (QAOA), and quantum neural networks, all of which combine parameterized quantum circuits (PQCs) with classical computation. While classical optimization is a common component, VQAs can incorporate diverse forms of classical processing, including large-scale post-processing, data analysis, and simulation (see table~\ref{tab:quantum-hpc-scenarios}). These algorithms exhibit loose to medium coupling between quantum and classical components.

Subspace quantum diagonalization (SQD) is another example of a PQC-based hybrid algorithm with substantial classical post-processing~\cite{kanno2023quantumselectedconfigurationinteractionclassical, shirakawa2025closedloopcalculationselectronicstructure, robledomoreno2024chemistry}. In SQD, the QPU prepares an approximate ground state via a PQC ansatz and samples it, producing measurement bitstrings that identify the most relevant configurations. The classical component then constructs a subspace energy matrix and solves the resulting eigenvalue problem to obtain a refined ground-state energy. This medium-coupled, task-parallel pattern can demand significant HPC resources: Shirakawa et al.~\cite{shirakawa2025closedloopcalculationselectronicstructure} demonstrate the workflow by coupling IBM's Heron QPU with up to 152,064 nodes of Fugaku for the classical post-processing. 

For Hamiltonian simulation, Mohseni et al.~\cite{mohseni2025buildquantumsupercomputerscaling} describe scaling approaches using distributed state vector simulation, tensor network methods, and circuit cutting to partition quantum workloads across multiple QPUs. This mode enables the gradual adoption of quantum computing within existing HPC applications and facilitates comparative studies between quantum and classical approaches.

The \emph{Quantum-about-HPC} mode focuses on integrating hybrid quantum and classical components into end-to-end applications and workflows (table~\ref{tab:quantum-hpc-scenarios}). In this mode, a quantum capability is commonly exposed as a workflow stage (often treated as a black-box service) that is embedded within application-specific data preparation, integration, and downstream analysis. Consequently, quantum and classical stages are frequently loosely coupled and communicate primarily through data transformations and intermediate artifacts. Representative examples include domain- and data-specific pre- and post-processing, such as encoding, loading, and converting data for machine learning~\cite{Kiwit2025}, and preconditioning quantum algorithms, e.\,g., warm-starting QAOA with a classical solution. In other cases, quantum outputs feed subsequent classical or quantum stages; for instance, samples produced by quantum generative models (e.\,g., QGANs, QCBMs) can drive further optimization or numerical simulation in application domains such as quantum chemistry and high-energy physics~\cite{Rehm:2824092}.

\section{Conceptual Middleware Architecture and State of the Art}
\label{sec:state-of-art}

This section first introduces a four-layer conceptual middleware architecture for quantum-HPC systems (section~\ref{sec:conceptual_architecture}), then surveys state-of-the-art quantum software and middleware systems and maps them to these layers (section~\ref{sec:sota}).

\subsection{Conceptual Middleware Architecture}
\label{sec:conceptual_architecture}

The quantum-HPC integration modes and application scenarios described above place complex requirements on resource management, from co-allocation for tightly-coupled tasks to effective support of heterogeneous, loosely-coupled tasks. Effective quantum-HPC resource management requires an adaptive, multi-level approach that leverages application-specific knowledge while accommodating quantum computing's unique characteristics~\cite{saurabh2023conceptual, rocco2025dynamic, BECK202411, elsharkawy2024integrationquantumacceleratorshpc}.

Addressing these gaps requires middleware that abstracts heterogeneous resource pools (CPUs, GPUs, QPUs) and supports diverse task-resource binding models, from co-allocation with fixed assignments for tightly-coupled tasks to late binding that enables load balancing and runtime adaptation for loosely-coupled workloads~\cite{10.1145/369028.369109}.

We realize these capabilities through a layered middleware architecture that decomposes resource management into four levels~\cite{Turilli_2019, 10.1145/3177851,6404423}, enabling multi-level and application-level scheduling where each layer handles a distinct aspect of quantum-HPC orchestration:

\emph{Workflow Layer (L4):} manages high-level scientific workflows and user interactions. This layer provides interfaces for workflow specification, manages dependencies between quantum and classical components, and coordinates long-running experiments that may span multiple resource allocations. The workflow layer abstracts the underlying complexity of hybrid execution while providing users with familiar interfaces for scientific computing.

\emph{Workload Layer (L3):} focuses on application-level resource management and scheduling decisions. This layer makes strategic resource allocation decisions, determines optimal parallelization strategies, and manages the mapping between logical application requirements and physical resource capabilities. The workload layer incorporates application-specific knowledge to optimize performance while respecting user-defined constraints and preferences. It receives workloads, i.e., a set of tasks, selects and allocates resources, partitions the workload across these resources, and binds tasks to resources.

\emph{Task Layer (L2):} handles the execution of individual computational tasks within the broader workload context. This layer manages task queuing, monitors execution progress, implements load balancing and fault tolerance mechanisms, and coordinates data movement between quantum and classical components. 

\emph{Resource Layer (L1):} encapsulates heterogeneous quantum and classical resources and schedules computational tasks to nodes, processors, and QPUs. Key challenges at this layer include integrating quantum resources and supporting tight coupling between quantum and classical tasks (e.\,g., for error correction and dynamic circuits), which requires co-allocation for low-latency interactions. While the traditional accelerator model allocates QPUs exclusively to single applications, QPU scarcity often demands multi-tenancy and non-exclusive access.

This layered decomposition separates strategic from operational concerns: upper layers (L4/L3) determine \emph{what} to execute and how to partition workloads across resources, incorporating application semantics such as circuit structure, fidelity requirements, and coupling patterns; lower layers (L2/L1) determine \emph{where} and \emph{when} to execute individual tasks, incorporating dynamic system state such as resource availability and execution feedback. Scheduling decisions are hierarchically delegated from the workflow layer to the resource layer, with each level refining the decisions of the level above.

\subsection{State of the Art}
\label{sec:sota}

\begin{figure}[t]
    \centering
    \includegraphics[width=0.65\columnwidth]{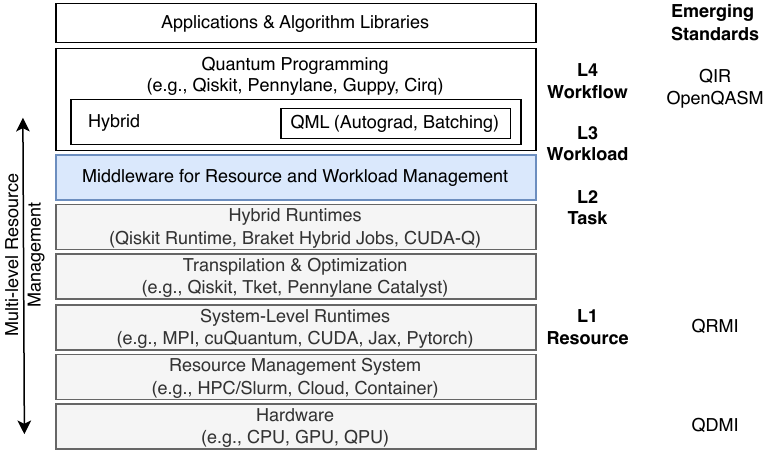}\upp\upp
    \caption{\textbf{Quantum Software Stack:} An overview of key components and layers, from quantum programming environments and hybrid runtimes to hardware resource management, including emerging standards.}
    \label{fig:quantum-software}
    \upp \upp
\end{figure}

We now survey existing quantum software and middleware systems, mapping them to the architectural layers defined above (see figure~\ref{fig:quantum-software}).

\subsubsection{Workflow Layer (L4)}

This layer focuses on workflow composition and orchestration. It provides well-defined programming abstractions for specifying quantum algorithms, as well as higher-level domain-specific libraries and workflow orchestration libraries.

\textit{Composition:} Quantum programming frameworks provide abstractions grounded in the quantum-circuit model to specify, transform, and execute programs. Core capabilities include circuit definition, optimization/transpilation, simulation, and execution on hardware backends.
Representative frameworks include Qiskit~\cite{javadiabhari2024quantumcomputingqiskit,Qiskit_OpenSource}, PennyLane~\cite{bergholm2022pennylaneautomaticdifferentiationhybrid}, Cirq~\cite{Cirq2024}, Qrisp~\cite{seidel2024qrispframeworkcompilablehighlevel}, Guppy~\cite{koch2025imperativequantumprogrammingownership}, Silq~\cite{10.1145/3385412.3386007}, and Q\#~\cite{Svore_2018}. Some frameworks specialize in specific domains, e.\,g., PennyLane for differentiable machine learning workflows, Qiskit Optimization~\cite{qiskit_optimization}, Qiskit Nature~\cite{qiskit_nature}, and OpenFermion~\cite{mcclean2019openfermionelectronicstructurepackage} for molecular systems.

Most programming frameworks provide integration with low-level quantum resources through pluggable backend providers and export circuits via intermediate representations, primarily OpenQASM~\cite{OpenQASM} and QIR~\cite{QIRAlliance}. For high-performance simulation, libraries such as cuQuantum~\cite{bayraktar2023cuquantum} provide GPU-accelerated state vector and tensor network backends that frameworks like CUDA-Q leverage. However, these capabilities often require manual resource management; a key challenge is orchestrating and scaling heterogeneous workloads comprising both tightly and loosely-coupled tasks.

\textit{Orchestration:} Higher-level workflow systems have emerged to connect coarse-grained application components and coordinate their execution. Workflow tools such as Prefect~\cite{narayanan2024orchestrating},  Covalent~\cite{covalent2023}, pyCOMPs/Qdislib~\cite{10.1145/3731599.3767547} include integrations for quantum libraries and backends (e.\,g., Prefect--Qiskit~\cite{prefect-qiskit}). Qiskit Functions~\cite{IBMQiskitServerless} provides managed, pre-built quantum workflows and function-style execution.

\subsubsection{Workload and Task Layer (L3, L2)}

Traditional HPC scheduling operates at the job level, making coarse-grained and static resource allocation decisions based primarily on user specifications and fairness policies. Quantum-HPC workloads require more sophisticated approaches that consider application characteristics and runtime dynamics, enabling adaptive task placement and resource binding. We survey the state of the art along three dimensions: \emph{execution runtimes} that provide infrastructure for quantum-classical workload management (distinguishing static and dynamic binding models), \emph{scheduling algorithms} that optimize resource selection based on application-specific and system-level information, and \emph{simulation tools} that enable scheduler development without consuming scarce quantum resources.

\paragraph{Execution Runtimes} 

Execution runtimes manage the binding of quantum-classical tasks to heterogeneous resources. We categorize them by the binding models introduced above: early-binding runtimes that co-allocate resources with fixed assignments at job or session start, and late-binding runtimes where task-resource mappings adapt at runtime.

\emph{Early-Binding Runtimes:} These systems support tightly-coupled, latency-sensitive scenarios such as variational algorithms with parameter updates, mid-circuit measurements, and error correction, where co-allocated resources remain fixed throughout execution.
Compilation frameworks are predominantly early-binding: XACC~\cite{xacc_2020}, QCOR~\cite{nguyen2020extendingcheterogeneousquantumclassical}, and CUDA-Q~\cite{CUDAQ} permit runtime choice of backend (e.\,g., at launch or via API), but once set, all kernels in that run execute on that target, i.e., there is no per-task reassignment or adaptive placement across backends. Cloud runtime services such as Qiskit Runtime~\cite{Qiskit_IBM_Runtime} and Braket Hybrid Jobs~\cite{AWSBraketJobs,braket-jobs-2021} bind classical and quantum resources via a  session mechanism enabling co-allocation with dedicated QPU access.

\emph{Late-Binding Runtimes:} These middleware systems support loosely-coupled scenarios requiring runtime adaptability. Representative systems include QFaaS~\cite{nguyen2024qfaas}, Tierkreis~\cite{https://doi.org/10.48550/arxiv.2211.02350}, Qibo~\cite{qibo_paper}, Quantum Framework~\cite{shehata2024frameworkintegratingquantumsimulation}, Q-IRIS~\cite{Miniskar:2025wcb}, Divi~\cite{qoro_divi}, and Orquestra~\cite{zapata2021orchestra}. Qiskit Serverless~\cite{ibm_quantum_serverless_2023,IBMQiskitServerless} manages hybrid workloads using Apache Ray~\cite{ray} for distributed execution, providing task decorators for resource mapping across CPUs, GPUs, and QPUs. Covalent~\cite{covalent_zenodo,covalent2023} models applications as DAGs with a dispatch service that maps tasks to user-defined executors via Dask~\cite{dask}; however, HPC integration remains experimental~\cite{covalent_hpc_plugin}.

\paragraph{Scheduling Algorithms} 

While the execution runtimes above provide infrastructure for workload management, effective quantum-HPC resource utilization additionally requires scheduling algorithms that incorporate dynamic system state, such as QPU calibration drift and noise characteristics, and application characteristics, e.\,g., workload partitioning trade-offs, into task placement decisions~\cite{10.1145/369028.369109}.

QFOR~\cite{nguyen2025qforfidelityawareorchestratorquantum} exemplifies prediction-based scheduling for QPU selection, using deep reinforcement learning with noise-aware estimators derived from calibration data and circuit properties to optimize execution time and fidelity. However, it is limited to per-task device selection and cannot orchestrate end-to-end hybrid workflows, multi-stage algorithms, or DAG-based execution. Qonductor~\cite{giortamis2025qonductor} extends this approach with hybrid resource estimation and many-to-many scheduling, using various optimization models trained on real execution data to balance fidelity and runtime. For simulation workloads, Maestro~\cite{maestro2025} proposes a model to automatically select the optimal simulation backends (state vector, tensor network, MPS, and GPU-accelerated) based on circuit structure and available hardware.

MILQ~\cite{10821104} jointly optimizes cut placement and subcircuit-to-resource mapping for circuit cutting workloads, achieving up to 25\% reduction in makespan; Qdislib~\cite{10.1145/3731599.3767547} similarly distributes subcircuits via PyCOMPSs across CPUs, GPUs, and QPUs. These systems illustrate how application characteristics, here circuit structure and cutting tradeoffs, can directly inform scheduling decisions.

\emph{Simulation:} Training and validating such schedulers requires extensive experimentation that is impractical on scarce, expensive quantum hardware. To address this challenge, simulation tools have emerged that model the dynamics of quantum-HPC infrastructure, enabling the development and evaluation of schedulers without consuming physical QPU resources. iQuantum~\cite{https://doi.org/10.1002/spe.3331} provides discrete-event simulation for quantum-HPC infrastructure, modeling QPU characteristics (qubit count, quantum volume, CLOPS, gate sets, topology) and supporting job scheduling with qubit requirements and topology constraints. HybridCloudSim~\cite{10.1145/3731599.3767548} extends this to hybrid scenarios, modeling iterative quantum-classical feedback loops under noise and resource constraints. However, neither fully captures HPC integration challenges such as coordinated scheduling across Slurm and quantum processors.

\emph{Discussion:} The execution runtimes surveyed above provide infrastructure for quantum-classical workload management but only offer limited application-level scheduling capabilities. Static runtimes fix task-resource assignments at submission, while dynamic runtimes (e.\,g., Qiskit Serverless, Covalent) offer execution flexibility but are not yet optimized for specific workload and application characteristics (e.\,g., circuit cutting). The scheduling algorithms (e.\,g., QFOR, Qonductor) advance prediction-based resource selection but remain limited to individual task-device mapping without end-to-end workflow orchestration. Pilot-Quantum (section~\ref{sec:pilot-quantum}) addresses the orchestration gap through the pilot abstraction, providing late binding and load balancing across heterogeneous resources. Q-Dreamer (section~\ref{sec:q-dreamer}) complements this with predictive workload optimization, determining \emph{what} to execute (e.\,g., optimal circuit cutting configurations) before Pilot-Quantum determines \emph{where} and \emph{when}.

\subsubsection{Resource Layer (L1)}

Standard interfaces and intermediate representations are emerging to provide vendor-agnostic backend access at this layer. The Quantum Device Management Interface (QDMI)~\cite{qdmi2025} abstracts device control and real-time hardware querying, while the Quantum Resource Management Interface (QRMI)~\cite{qrmi2025} provides APIs for resource acquisition and task execution; formal specifications such as DIN SPEC 91520~\cite{din-spec} define interoperability requirements. Intermediate representations, primarily OpenQASM~\cite{Cross_2022} and QIR~\cite{QIRAlliance, lubinski2022advancinghybridquantumclassicalcomputation}, enable transpilers and compilers (e.\,g., Tket~\cite{tket}, BQSKit~\cite{osti_1785933}, Staq~\cite{staq}) to transform circuits into hardware-optimized forms. Despite these efforts, HPC workload managers have only begun to integrate quantum resources; for example, QRMI's Slurm SPANK plugin~\cite{qrmi2025} represents an early effort to expose QPUs as schedulable entities within existing job schedulers.

\section{Understanding Applications: Execution Motifs and Quantum Mini-Apps} 
\label{sec:motifs}

Systematically evaluating and comparing middleware systems requires a structured characterization of quantum-HPC workloads. We address this through execution motifs and quantum mini-apps~\cite{saurabh2024quantum, saurabh2025compositional}: motifs capture recurring patterns of quantum-classical interaction that operationalize the integration modes introduced in section~\ref{sec:background}, from tightly-coupled basic motifs in HPC-for-Quantum scenarios to loosely-coupled compositional motifs in Quantum-about-HPC workflows. Mini-apps are compact implementations that instantiate these patterns, serving as practical benchmarks for evaluating middleware performance.

\begin{figure}[t]
  \centering
  \includegraphics[width=\textwidth]{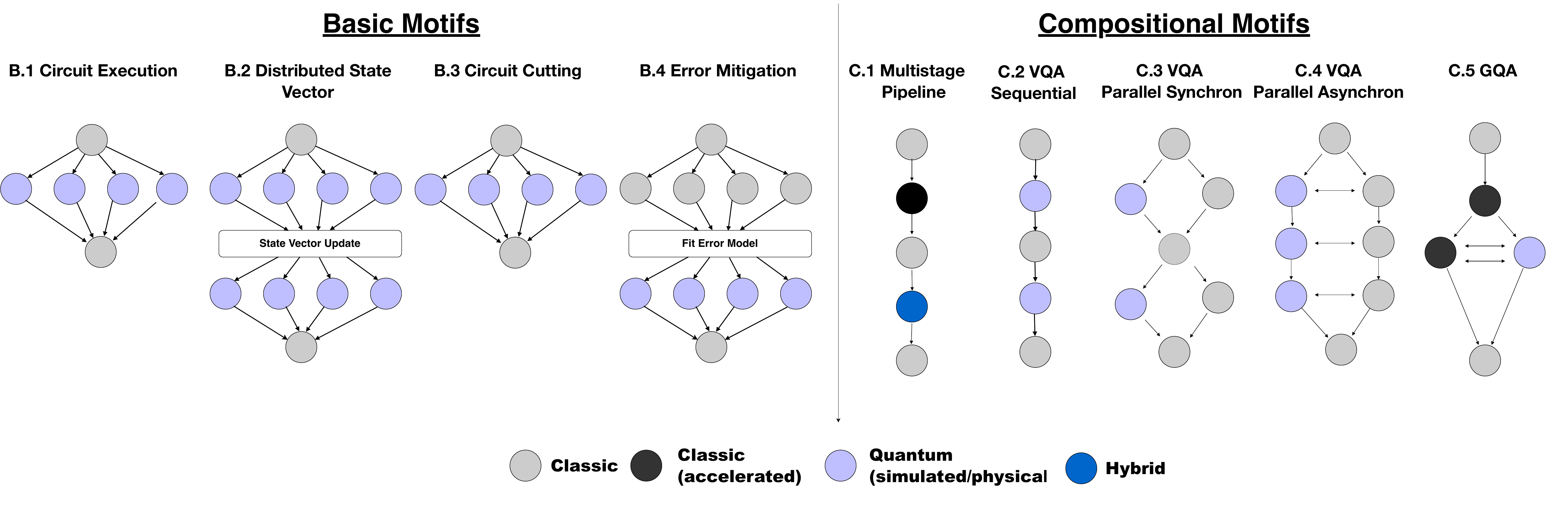}
  \caption{\textbf{Basic and Compositional Execution Motifs for Quantum-HPC Workflows:} Basic motifs represent fundamental patterns of quantum computation and classical-quantum interaction, including circuit execution, distributed simulation, circuit cutting, and error mitigation. These motifs are characterized by their coupling intensity (tight vs. loose) and interaction patterns (concurrent vs. sequential). Compositional motifs orchestrate multiple basic motifs to create real-world applications, such as multi-stage pipelines, parallel VQAs with synchronous or asynchronous coordination, and generative quantum algorithms that combine classical machine learning with quantum circuit execution.}
  \label{fig:motifs}
\end{figure}

\subsection{Execution Motifs}

Motifs~\cite{saurabh2024quantum} are recurring high-level patterns of interaction and coupling between quantum and classical components within a hybrid workflow. While the quantum-HPC integration modes in section~\ref{sec:background} describe the architectural relationship between quantum and HPC systems, motifs describe how tasks interact at runtime. Motifs are characterized along two dimensions: interaction patterns (e.\,g., concurrent vs. sequential task execution) and coupling intensity, which ranges from tight (within QPU coherence time) to loose (end-to-end workflow integration). Importantly, complex applications often exhibit heterogeneous coupling and interaction patterns that vary across stages: a single workflow may combine tightly coupled error correction with loosely coupled classical optimization, posing significant challenges for middleware that must dynamically adapt scheduling and resource allocation strategies.

Table~\ref{tab:execution_motifs} summarizes a representative, though not exhaustive, set of motifs derived from the application scenarios above; additional motifs may emerge as new domains and integration strategies develop.
We categorize motifs as either basic or compositional, where basic motifs represent fundamental patterns and compositional motifs orchestrate multiple basic patterns into more complex applications (figure~\ref{fig:motifs})~\cite{saurabh2024quantum,saurabh2025compositional}.

\begin{table*}[t]
  \centering
  \caption{{\bf Quantum-HPC Execution Motifs:} {\normalfont Characterization of quantum-classical interaction patterns organized by complexity: Basic Motifs represent fundamental patterns, while Compositional Motifs orchestrate multiple basic patterns into complex workflows~\cite{saurabh2024quantum,saurabh2025compositional}.}}
  \tiny
  \label{tab:execution_motifs}
  \setlength{\tabcolsep}{1pt}
  \renewcommand{\arraystretch}{1.3}
  
  \begin{tabular}{|>{\raggedright\arraybackslash}p{1.9cm}|>{\raggedright\arraybackslash}p{3.2cm}|>{\raggedright\arraybackslash}p{3cm}|>{\raggedright\arraybackslash}p{2.3cm}|>{\raggedright\arraybackslash}p{3.2cm}|}
  \hline
  \textbf{Motif} & \textbf{Interaction} & \textbf{Coupling} & \textbf{Example} & \textbf{Middleware Requirements}\\
  \hline
  \multicolumn{5}{|c|}{\cellcolor{gray!20}\textbf{Basic Motifs}}\\
  \hline
  B.1 Circuit Execution & Concurrent computation, e.\,g., for different measurements or parameters & Loosely-coupled, homogeneous tasks & Qiskit Estimator/Sampler, Qiskit Aer with Dask~\cite{QiskitAerParallel} & Improved workload management across heterogeneous resources\\
  \hline
  B.2 Distributed State Vector Simulation & Concurrent computation of state vector updates \& its synchronization across all tasks & Tightly-coupled static and homogeneous tasks & Large-scale quantum simulations, e.\,g., cuQuantum~\cite{cuquantum} & Integration with low-level MPI runtime\\
  \hline
  B.3 Circuit Cutting & Concurrent execution \& reconstruction of tasks across QPUs & Medium coupling, task heterogeneity depends on circuit type \& cutting algorithm & Qiskit Circuit Knitting Toolbox~\cite{qiskit-addon-cutting}, Pennylane Circuit Cutting~\cite{PennyLaneQCut} & Optimal partitioning \& placements of tasks/cuts on available (simulated) QPUs for balanced workload\\
  \hline
  B.4 Error Mitigation & Loosely-coupled execution of multiple circuit variants on PEs \& results aggregation & Loosely coupled & Qiskit Runtime Error Mitigation~\cite{IBMQuantumConfigureErrorMitigation} & Allocation \& adaptation of error mitigation to right mix of classical \& quantum resources\\
  \hline
  \multicolumn{5}{|c|}{\cellcolor{gray!20}\textbf{Compositional Motifs}}\\
  \hline
  C.1 Multistage Pipelines & Encapsulated stages \& its contained transitions with control \& data flow & Heterogeneous \& varying resource demands between stages & QML workflow including data encoding step~\cite{Mantha_Pilot-Quantum_2024} & Optimized resource estimation for pipeline/individual stages, e.\,g., for dynamic resource pool\\
  \hline
  C.2 Variational Quantum Algorithms (sequential) & Concurrent execution of classical \& quantum components & Coupling outside coherence window for heterogeneous tasks, e.\,g., interleaved ML \& QC parts sharing GPU & QuGen~\cite{QutacQuantumQuGen}, VQE, QAOA, QIRO~\cite{Fin_gar_2024}  & Collocate quantum \& classical resources\\
  \hline
  C.3 Synchronous Parallel VQA & Parallel circuit evaluations with barrier synchronization & Medium coupling; synchronization required at optimization iteration boundaries & pQAOA~\cite{cattelan2023parallelcircuitimplementationvariational}, distributed QAOA~\cite{kim2025distributedquantumapproximateoptimization}, VQA-IS~\cite{vqa_is} & Coordinated scheduling across QPUs/simulators; barrier synchronization; load balancing for parallel circuit batches\\
  \hline
  C.4 Asynchronous Parallel VQA & Independent optimization tasks with occasional synchronization & Loose coupling; tasks executed independently& EQC~\cite{stein2021eqcensembledquantum} & Asynchronous task management; fault tolerance; convergence detection without global barriers\\
  \hline
  C.5 GQA & Synchronous \& asynchronous exchange between inference \& circuit execution & Tight coupling (GPU-accelerated training); loose-coupling (inference \& circuit execution) & GPT-QE~\cite{nakaji2024generative}, GQE~\cite{minami2025generativequantumcombinatorialoptimization} & Co-scheduling AI frameworks \& QPU runtimes; resource allocation adjustment based on varying workload (feedback parameters)\\
  \hline
  \end{tabular}
\end{table*}

\emph{Basic Execution Motifs:} We identify four basic motifs that are pervasive:
(B.1)~\emph{Circuit Execution} encompasses parametric sweeps and ensemble processing with loosely-coupled, homogeneous tasks;
(B.2)~\emph{Distributed Simulation} addresses state vector, tensor-network, and stabilizer simulations using tightly-coupled MPI/GPU parallelism;
(B.3)~\emph{Circuit Cutting} decomposes large circuits into fragments for parallel execution on resource-constrained devices; and
(B.4)~\emph{Error Mitigation} executes multiple circuit variants for noise reduction techniques such as zero-noise extrapolation and probabilistic error cancellation.

\emph{Compositional Execution Motifs:} We identify five compositional motifs that capture the heterogeneous, multi-stage nature of real-world quantum-HPC applications:
(C.1)~\emph{Multi-Stage Pipelines} combine quantum and classical processing stages with varying resource demands, e.\,g., QML training workflows;
(C.2)~\emph{Sequential VQAs} capture iterative optimization with repeated cycles of circuit execution and classical parameter updates;
(C.3)~\emph{Synchronous Parallel VQAs} extend C.2 with multiple optimization or other classical tasks (e.\,g., classical solvers) sharing information at regular synchronization points;
(C.4)~\emph{Asynchronous Parallel VQAs} relax synchronization for flexible resource allocation, with independent tasks and periodic information sharing; and
(C.5)~\emph{Generative Quantum Algorithms} leverage classical generative AI models to design quantum circuits, requiring coordination between GPU-based training and quantum execution~\cite{nakaji2024generative,minami2025generativequantumcombinatorialoptimization}.

\subsection{Quantum Mini-Apps}

While motifs describe workload patterns, quantum mini-apps~\cite{saurabh2024quantum, QuantumMiniApp} instantiate them as concrete, executable benchmarks. Mini-apps are representative prototypes that capture the essential performance characteristics of specific algorithmic kernels and their middleware requirements. They serve three purposes: characterizing heterogeneous hardware within application contexts, providing standardized workloads for middleware evaluation, and supporting hardware-software co-design without requiring a complete stack redesign. The current suite includes quantum simulation mini-apps (circuit execution, circuit cutting, and distributed state vector simulation) and quantum machine learning mini-apps (classification, data compression, and variational training).

\subsection{Discussion}

The motif analysis reveals two complementary requirements. First, motifs range from loosely-coupled patterns (circuit execution, error mitigation) to tightly-coupled patterns (distributed simulation), often requiring access to various computational resources like GPUs, QPUs, and CPUs; this demands a unified resource abstraction supporting diverse task-resource binding models, from co-allocation to late binding. Second, no single scheduling strategy suffices: loosely-coupled motifs benefit from application-level task scheduling (L4/L3), while tightly-coupled motifs require co-scheduling at the resource layer (L2/L1). Bridging this gap between the upper layers (L3/L4) and lower layers (L1/L2) is the core middleware role. Pilot-Quantum (section~\ref{sec:pilot-quantum}) addresses the abstraction challenge through the pilot model, providing a uniform interface with late-binding task-resource mapping; Q-Dreamer (section~\ref{sec:q-dreamer}) complements this with workload partitioning recommendation informed by application semantics.

\section{Pilot-Quantum: A Middleware for Hybrid Quantum-HPC Applications} 
\label{sec:pilot-quantum}

Pilot-Quantum operates at the workload and task layers (L3/L2) of our conceptual architecture, addressing the two challenges identified by the motif analysis: infrastructure heterogeneity and application heterogeneity. It builds on the pilot abstraction~\cite{10.1145/3177851, 6404423}, which enables flexible resource management and adaptive application-level scheduling across heterogeneous quantum-HPC environments. Q-Dreamer (section~\ref{sec:q-dreamer}) complements Pilot-Quantum with workload partitioning recommendations informed by application semantics.

\emph{Managing heterogeneous quantum-HPC infrastructure:} Pilot-Quantum provides a unified orchestration layer across diverse classical resources (CPUs, GPUs, cloud instances), quantum backends (simulators and QPUs from multiple vendors), and quantum software frameworks (Qiskit, PennyLane, and cuQuantum). An extensible plugin mechanism supports integration of new backends and execution engines, enabling workloads to span heterogeneous runtimes while preserving framework- and backend-specific optimizations (e.\,g., via framework- and runtime-specific plugins).

\emph{Enabling application-level scheduling for heterogeneous workflows:} Pilot-Quantum exposes task and workload abstractions that support application-level scheduling based on application semantics (e.\,g., circuit structure, fidelity requirements, and quantum-classical coupling patterns) and dynamic system state (e.\,g., resource availability and observed execution behavior). This supports both basic and compositional motifs, including tightly-coupled and loosely-coupled execution patterns, and enables dynamic resource allocation and late binding as workflow demands change across stages.

\subsection{Pilot Abstraction}

In the pilot model~\cite{10.1145/3177851, 6404423}, a pilot (L2) is a placeholder job that acquires resources from the system scheduler (L1), maintains them for a specified duration, and handles runtime task-resource binding within those resources. The \emph{Pilot-Manager} (L3) orchestrates one or more pilot agents, making strategic decisions about pilot placement and task distribution based on application characteristics and system state. This creates a \emph{multi-level scheduling} architecture: the system scheduler (e.\,g., Slurm) handles coarse-grained resource allocation at L1, while the Pilot-Manager and pilots handle fine-grained, application-aware task placement at L3/L2. By decoupling resource acquisition from task execution, the pilot abstraction enables the late-binding capability identified as critical for handling infrastructure and application heterogeneity, enabling application-level scheduling that incorporates application semantics, such as circuit structure and coupling patterns, into placement decisions at runtime.

This multi-level architecture is particularly important for quantum-HPC workloads, which are both dynamic and require adaptive scheduling. Quantum-HPC infrastructure and applications are dynamic: resource demands shift across workflow stages (e.\,g., a QML pipeline may require GPU-heavy data encoding, followed by QPU-intensive circuit execution, and finally CPU-bound classical post-processing), and infrastructure availability changes. Thus, scheduling must be adaptive and able to respond to feedback during execution such as changing resource availability or application-internal triggers (e.\,g., convergence criteria in variational algorithms). Static job-level allocation through system schedulers cannot accommodate either dimension; the pilot abstraction allows the Pilot-Manager to redistribute tasks across pilots or allocate/deallocate new pilots as conditions evolve. In practice, this enables efficient batching and distribution of quantum circuits and subcircuits across heterogeneous backends, maximizing QPU utilization while adapting to runtime conditions.

\subsection{Architecture}

\begin{figure}[t]
  \centering
  \includegraphics[width=0.7\textwidth]{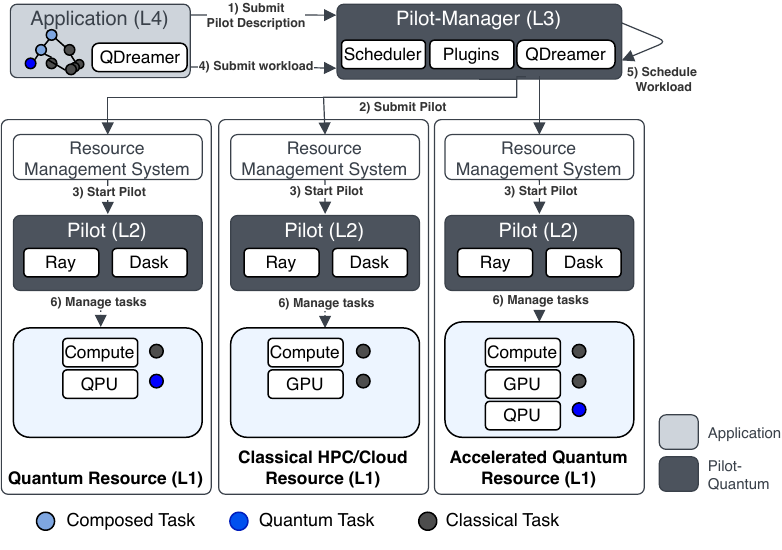}\upp \upp
  \caption{\textbf{Pilot-Quantum Architecture:} The system core is a \emph{Pilot-Manager} that orchestrates and manages resources through pilots across both classical and quantum infrastructures, such as QPUs, GPUs, and CPUs. Pilots are responsible for reserving resources and managing task execution.}
  \label{fig:quantum-pilot-arch}
  \upp \upp
\end{figure}

Figure~\ref{fig:quantum-pilot-arch} shows Pilot-Quantum's architecture. For completeness, the figure also shows Q-Dreamer, which provides performance-model-assisted recommendations for workload partitioning and resource allocation (e.\,g., for circuit cutting workloads); see section~\ref{sec:q-dreamer}. The Pilot-Manager provides extensible resource plugins to integrate various backends. Its global scheduler makes strategic decisions about pilot placement and task distribution based on application characteristics and system state. The following describes how Pilot-Quantum implements and integrates each layer of our conceptual architecture respectively.

\emph{L4 Workflow Layer:}
Pilot-Quantum provides two abstractions: (i) resource management, enabling applications to allocate appropriate classical and quantum resources, and (ii) workload and task management, facilitating efficient workload execution on these resources. For (i), applications must define required resources in a \texttt{pilot\_description} and submit it to Pilot-Quantum, which then acquires the resources.

For (ii), applications must decompose their problems into independently executable tasks, collectively forming the workload. Tasks can encapsulate units of quantum computations (e.\,g., circuit executions) and classical computations (e.\,g., pre- and post-processing). While early quantum applications require high control over problem decomposition and workload, Pilot-Quantum is expected to interface with higher-level workflow systems and application libraries. 

Tasks are defined using a \texttt{task\_description} and submitted via the \texttt{submit\_task} API. Pilot-Quantum also integrates with native abstractions of different frameworks, e.\,g.,  Dask and Ray. Tasks can be assigned (i) directly to a specific pilot or (ii) to a group of pilots. For (ii), the pilot manager assigns tasks to pilots, optimizing throughput by load balancing across all pilots. Please refer to \cite{pilot-quantum-api-usage} for API details.

\emph{L3/L2 Workload and Task Layers:}
After receiving the \texttt{pilot\_description}, the Pilot-Manager allocates requested resources via the local resource management system (\emph{steps 1-3} in figure~\ref{fig:quantum-pilot-arch}). In HPC environments, a placeholder job that starts the Pilot-Agent is queued via Slurm. Once active, the Pilot-Agent manages the allocated resources through a local task queue and a monitor for tracking performance and resource availability. Tasks are dispatched through pluggable execution engines matched to workload characteristics: Dask~\cite{pilot-streaming} for data-parallel and streaming tasks, Ray~\cite{ray} for GPU-accelerated workloads such as quantum simulations, and MPI/CUDA via Slurm's \texttt{srun} for tightly-coupled tasks such as distributed state vector simulations. Quantum hardware access is handled via classical tasks that utilize framework-specific providers (e.\,g., Qiskit or PennyLane plugins).

For quantum hardware, Pilot-Quantum supports three distinct QPU access modes~\cite{elsharkawy2023integration}: (i)~dedicated node-attached QPUs with exclusive access during pilot lifetime, enabling tight coupling with co-located classical computation; (ii)~shared remote QPUs accessed through cloud APIs with multi-tenant queuing; and (iii)~session-based access, where cloud providers offer mechanisms (e.\,g., Qiskit Runtime Sessions, Amazon Braket Jobs) that provide temporary dedicated access with co-located classical execution. In shared and session-based modes, Pilot-Quantum cannot reserve QPU time directly; instead, the late-binding model enables dynamic adaptation to variable queue times, redirecting tasks between backends based on real-time availability.

\emph{L1 Resource Layer:}
Pilot-Quantum abstracts heterogeneous resources through its plugin mechanism for various runtimes and can be integrated with resource-specific tools, e.\,g., QPU-specific transpilers.

\section{Q-Dreamer Toolkit: Enabling Application-Level Workload Management} \label{sec:q-dreamer}

Our conceptual architecture and motif analysis identified the need for applications to make informed decisions about workload partitioning and resource allocation. While Pilot-Quantum provides the orchestration infrastructure that determines \emph{where} and \emph{when} to execute tasks, it does not determine optimal workload configurations. Q-Dreamer addresses this gap with reusable building blocks for resource detection and workload analysis, generating performance-model-assisted recommendations for workload partitioning and task granularity that Pilot-Quantum then executes.  We describe the Q-Dreamer architecture below and demonstrate its capabilities through the Circuit Cutting Resource Optimizer, which determines optimal partitioning strategies for large quantum circuits. Q-Dreamer is available as open-source software as part of the quantum mini-apps framework~\cite{radical_qminiapps_qdreamer_2026}.

\begin{figure}[t]
    \centering
    \includegraphics[width=0.6\columnwidth]{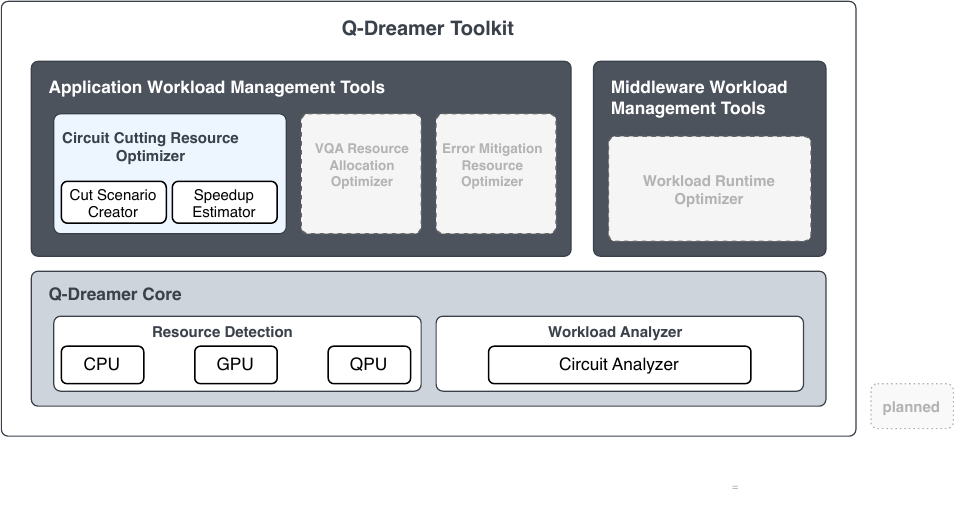}\upp\upp
    \caption{\textbf{Q-Dreamer Toolkit Architecture:} The Q-Dreamer framework consists of two main layers. The \emph{Q-Dreamer Core} layer provides re-usable building blocks for resource detection and workload analysis. The \emph{Workload Management Tools} layer provides application- and workload-specific tools to support scheduling decisions based on the Q-Dreamer core.}
    \label{fig:qdreamer-arch}
    \upp\upp
\end{figure}

\subsection{Q-Dreamer Architecture}

Figure~\ref{fig:qdreamer-arch} illustrates the Q-Dreamer architecture. Q-Dreamer Core provides foundational capabilities that two categories of tools build upon: Application Workload Management tools (L4/L3) generate workload configurations before execution, while Middleware Workload Management tools (L3/L2) will optimize task placement at runtime. The Circuit Cutting Resource Optimizer (CCRO) is the first Application Workload Management tool implemented on this foundation; Middleware Workload Management tools are planned as future work.

\emph{Q-Dreamer Core:} The core layer provides two capabilities. The \emph{Resource Detection} component characterizes available computing resources, including GPU device specifications, CPU core counts, and memory capacity, to inform resource-aware optimization decisions. The \emph{Workload Analyzer} extracts circuit characteristics, such as qubit count, circuit depth, gate composition, and entanglement patterns, that determine resource requirements and execution time estimates. Together, these components enable Q-Dreamer to match workload characteristics with available resources.

\emph{Application Workload Management Tools (L4/L3):} These tools combine the core capabilities with application-specific models to generate workload configurations before execution. The CCRO tool recommends partitioning strategies that balance subcircuit sizes against sampling overhead given the detected resource capacity (see section~\ref{sec:cc_resource_optimizer}).

\subsection{Circuit Cutting Resource Optimizer (CCRO)}
\label{sec:cc_resource_optimizer}

As introduced in the circuit cutting motif (B.3, section~\ref{sec:motifs}), circuit cutting~\cite{Piveteau_2024,Vazquez_ibm_circuit_cutting_2qpus} decomposes large circuits into smaller subcircuits that can be executed independently and recombined to reconstruct the original outcome. While this enables execution beyond the qubit limits of individual QPUs, the required number of subcircuit executions grows exponentially with the number of cuts~\cite{qiskit_circuit_cutting_sampling_overhead_table}; for example, a single CNOT gate cut incurs a sampling overhead factor of $9$~\cite{Martiel_2021,schmitt2024cuttingcircuitsmultipletwoqubit}.

Beyond this inherent overhead, circuit cutting workloads are complex to manage due to several reasons: (i) Complex resource demands: while the circuit partitioning and reconstruction stages are CPU-intensive, subcircuit execution may target CPUs, GPUs, or QPUs depending on circuit characteristics, with smaller subcircuits efficiently simulated classically and larger subcircuits potentially benefiting from quantum execution; (ii) Imbalanced workload: Circuit cutting tools such as Qiskit Circuit Cutting Addon~\cite{qiskit-addon-cutting} frequently generate imbalanced task distributions, where subcircuits of varying computational complexity lead to resource stragglers that degrade parallel efficiency and leave compute resources idle.

These characteristics motivate the need for intelligent resource optimization tools. The Circuit Cutting Resource Optimizer addresses the following problem: given a resource allocation (e.\,g., as defined by the currently active pilots managing CPUs, GPUs, and QPUs), determine the optimal number of cuts that maximizes execution speedup. Q-Dreamer solves this by analyzing the tradeoffs between the number of cuts, subcircuit sizes, sampling overhead, reconstruction costs, and parallel task distribution across available resources. The optimizer recommends an optimal subcircuit size for a given circuit and resource configuration.

\emph{Speedup Estimation Model:}
To estimate the speedup $S$ for a given circuit cutting scenario, the CCRO uses a calibrated speedup model. Given a quantum circuit $Q$ with $n$ qubits and $W$ available workers (e.\,g., CPU cores or GPUs), the model approximates speedup as the ratio between the full-circuit simulation cost $C_{\text{full}}$ and the parallelized circuit cutting cost $C_{\text{cut}}$: $S = \tfrac{C_{\text{full}}(n)}{C_{\text{cut}}(n_{\text{sub}}, k, W)}$, where $n_{\text{sub}}$ is the subcircuit size and $k$ is the number of cuts.\footnote{For simplicity, we omit constant factors in the cost expressions; these are captured by the calibration parameters $\eta_{\max}$ and $p$ when fitted to empirical data.} The full-circuit cost scales exponentially with qubit count, reflecting the $2^n$ state vector size: $C_{\text{full}}(n) \;=\; 2^{n}.$

\emph{Cutting Cost:}
Circuit cutting decomposes the computation into $N_{\text{tasks}} = 9^{k}$ subexperiments. This scaling arises because each cut introduces a complete Pauli-basis expansion: three measurement bases and three state preparations, yielding $3 \cdot 3=9$ configurations per cut and $9^k$ total subexperiments for $k$ cuts~\cite{mitarai2021constructing}. Given $W$ workers, tasks are executed in \emph{rounds}: $R \;=\; \left\lceil \frac{9^{k}}{W} \right\rceil.$
For fixed $W$, each additional cut multiplies the number of rounds by approximately $9$, making $R$ exponential in $k$.

To account for parallel inefficiencies (scheduling, synchronization, load imbalance), Q-Dreamer models efficiency as a power-law decay in the number of rounds:
$\eta(R) = \tfrac{\eta_{\max}}{R^{p}}$,
where $\eta_{\max}\in(0,1]$ is the peak efficiency achievable with minimal rounds (capturing backend-specific overheads such as task dispatch and memory management), and $p \ge 0$ is the decay exponent controlling how rapidly efficiency degrades as $R$ increases (capturing synchronization and load-balancing penalties across rounds). The cutting cost combines subcircuit work, rounds, and efficiency:
\begin{equation}
  C_{\text{cut}}(n_{\text{sub}},k,W)
  \;=\;
  \underbrace{2^{n_{\text{sub}}}}_{\text{Subcircuit Work}}
  \cdot
  \underbrace{R(W,k)}_{\text{Rounds}}
  \cdot
  \underbrace{\frac{1}{\eta(R)}}_{\text{Efficiency Penalty}}.
\end{equation}

Combining the above yields the following speedup estimation (for $k>0$):
\begin{equation}
S \;=\; 2^{\,n-n_{\text{sub}}} \cdot \frac{\eta(R)}{R}
\end{equation}

\emph{Calibration:}
The parameters $\eta_{\max}$ and $p$ are calibrated for each backend (e.\,g., CPU, H100, B200) using least-squares regression of experimental measurements. Given observed speedups from calibration experiments with varying circuit sizes, subcircuit sizes, cut counts, and worker counts, Q-Dreamer first derives the implied efficiency for each measurement from the speedup formula. It then fits the power-law efficiency model to obtain $\eta_{\max}$ and the decay exponent $p$.

\emph{Practical Considerations:}
In practice, the decision to apply circuit cutting is determined by hardware limits. Statevector simulation stores $2^n$ complex amplitudes, imposing hard memory limits on feasible circuit sizes. When $n$ exceeds the resource's memory capacity, cutting becomes necessary rather than optional. Conversely, the overhead of cutting is only beneficial when the circuit size approaches hardware limits; for smaller circuits, direct simulation outperforms the costs of partitioning and reconstruction. Thus, a backend-specific memory threshold determines whether circuit cutting is used; below this threshold, direct simulation always outperforms it.

\emph{Limitations:}
The round-based performance model assumes uniform task complexity; in practice, subcircuits vary in computational cost, leading to straggler effects in which faster tasks wait for slower ones. As $k$ increases, the exponential growth of $9^k$ subexperiments amplifies these inefficiencies. Additionally, the calibrated parameters $\eta_{\max}$ and $p$ are workload- and hardware-dependent; a single calibration may not generalize across all factors.

\section{Evaluation and Results}
\label{sec:eval}

This section evaluates Pilot-Quantum and Q-Dreamer using the Quantum Mini-App framework~\cite{saurabh2024quantum, QuantumMiniApp}, which instantiates the execution motifs introduced in section~\ref{sec:motifs}. We evaluate four key motifs: (i)~circuit execution (B.1) across QPUs and simulators, (ii)~distributed simulation (B.2) for state vector scaling, (iii)~multi-stage pipelines (C.1) for quantum machine learning workflows, and (iv)~circuit cutting (B.3) with Q-Dreamer's workload optimization.

\emph{Experimental Setup:} We performed experiments on two platforms. \emph{Perlmutter} is a heterogeneous system based on the HPE Cray Shasta platform with \num{3072} CPU-only and \num{1792} GPU-accelerated nodes. GPU nodes comprise \num{64} CPU cores and \num{4} \texttt{A100} GPUs with \SI{40}{\giga\byte} or \SI{80}{\giga\byte} of GPU memory; CPU nodes comprise \num{128} cores with \num{2} CPUs and \SI{512}{\giga\byte} of memory. Experiments were orchestrated from a dedicated cluster node allocated via Slurm or Pilot-Quantum. Additionally, we used \emph{NVIDIA DGX} systems equipped with H100 GPUs (\SI{80}{\giga\byte} memory) and B200 GPUs (\SI{180}{\giga\byte} memory) for circuit-cutting experiments that required larger GPU memory capacity.

Our software stack includes Pilot-Quantum with Slurm and Ray plugins for resource orchestration. For quantum circuit execution and simulation, we use Qiskit (Aer v0.15.1, IonQ provider, Runtime library). For distributed state vector simulation, we employ PennyLane's \texttt{lightning.gpu} (v0.41.0) with cuQuantum for GPU-accelerated simulation. For QML workflows, we use PennyLane's \texttt{default.qubit} with JAX interface, leveraging JIT compilation and vmap optimization. Ray handles distributed task execution. Quantum backends include IBM Eagle QPU accessed via IBM cloud, IonQ cloud simulators, and Qiskit Aer simulators (CPU and GPU). 

For distributed state vector simulations, we used \num{64} nodes and \num{256} A100 GPUs. Circuit cutting experiments used Qiskit's circuit-cutting addon with EfficientSU2 ansatz, running on CPUs (up to \num{224} workers) and NVIDIA DGX B200/H100 GPUs, with \num{4096} shots per circuit and chunk-based memory management (\texttt{blocking\_enable=True}, \texttt{blocking\_qubits=31}, \texttt{batched\_shots\_gpu=True}).

\subsection{Pilot-Quantum Characterization}

\begin{table}[t]
  \centering
  \caption{Pilot-Quantum Scheduling Overhead: Average Throughput and Duration for Zero-Compute Tasks}
  \label{tbl:avg_throughput_duration}
  \upp 
  \resizebox{0.3\columnwidth}{!}
  {
  \begin{tabular}{|c|c|c|}
  \hline
  \emph{Tasks} & \emph{Runtime \SI{}{[\second]}} & \emph{Throughput \SI{}{[tasks/\second]}}
  \\
  \hline
  \num{256} & \num{0.6} & \num{418.1} \\
  \hline
  \num{512}  & \num{1.3} & \num{406.5} \\ 
  \hline
  \num{1024} & \num{2.6} & \num{397.6} \\
  \hline
  \num{2048}  & \num{5.8} & \num{358.6} \\ 
  \hline
  \num{4096} & \num{11.4} & \num{358.4} \\
  \hline
  \num{8192}  & \num{27.4} & \num{298.2} \\ 
  \hline
  \end{tabular}
  }
  \end{table}

Table \ref{tbl:avg_throughput_duration} shows Pilot-Quantum scheduling overhead analysis for \num{256} to \num{8192} zero-compute tasks (\texttt{/bin/sleep 0}) across \num{4} Perlmutter CPU nodes (\num{1024} threads). These tasks carry no computational payload, so the measured throughput of \num{298}--\num{418}~tasks/second reflects pure scheduling and task-dispatch overhead, independent of application workload. This constitutes an upper bound on achievable throughput; in practice, memory pressure, GPU contention, or QPU queue delays will reduce effective throughput. On average, \SI{37}{\second} were required for setting up the pilot. As task counts increased, the Ray engine exhibited significant object management overhead, resulting in reduced throughput \cite{ray_overhead}. Since circuit execution times range from seconds to minutes (section~\ref{sec:circuit_execution}), the per-task dispatch cost ($<\SI{4}{\milli\second}$) is negligible relative to actual workload.

\subsection{Circuit Simulation Mini-App}
\label{sec:circuit_simulation}

We extend the quantum simulation mini-app~\cite{saurabh2024quantum} by adding and evaluating different cloud backends (incl. physical quantum hardware)  (section~\ref{sec:circuit_execution}), an implementation
of distributed state vector simulation (section~\ref{sec:dist_state_vector}) and circuit cutting (section~\ref{sec:circuit_cutting}).

\subsubsection{Circuit Execution:}
\label{sec:circuit_execution}

\begin{figure}[t]
  \centering
  \resizebox{0.8\textwidth}{!}{
  \includegraphics{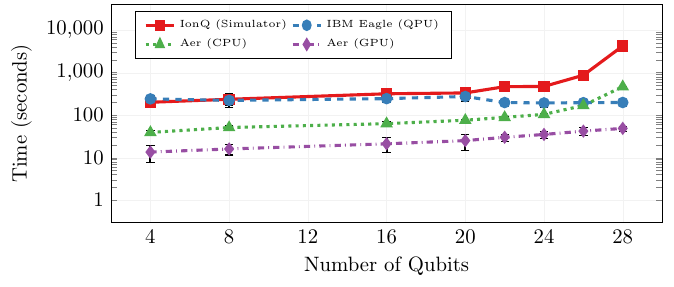}}
  \up

    \caption{\textbf{Circuit Execution on IonQ, IBM Eagle and Qiskit Aer (CPU and GPU):}
  Comparing end-to-end execution times for batches of random circuits (2 to 28 qubits) across IonQ Quantum Cloud, IBM Eagle, and Qiskit Aer simulators on CPU and GPU. For IonQ and Aer, we execute 1024 circuits (8 random circuits per qubit configuration). For IBM Eagle, we execute 8 circuits per qubit configuration and scale the mean runtime to the equivalent 1024-circuit workload for comparability. The results show exponential runtime growth with qubit count for cloud-based execution and CPU simulation, while GPU-accelerated simulation achieves substantially lower runtimes.}
  \label{fig:circuit_execution_ionq_pq_scaling} 
  \upp \upp
\end{figure}

We evaluated various quantum backends, including (i) in-process simulators using Qiskit Aer~\cite{QiskitAer}, (ii) IonQ cloud simulators accessed via Qiskit IonQ provider~\cite{ionq_quantum_cloud}, and (iii) IBM's Eagle QPU via IBM cloud using Qiskit's Runtime library. We executed \num{1024} random circuits across backends ranging from \num{2} to \num{28} qubits.

Figure~\ref{fig:circuit_execution_ionq_pq_scaling} shows circuit execution results. For the IonQ simulator, we observe that compute time increases exponentially with qubit count, from \SI{175.8}{\second} for \num{2} qubits to \SI{4260}{\second} for \num{28} qubits. This aligns with the expected exponential scaling of classical quantum circuit simulation due to exponential growth in the size of the state space. Qiskit Aer (CPU) backend showed similar scaling but lower execution times than IonQ, with \num{28} qubits taking \SI{469.7}{\second}. 
Aer (GPU) outperforms IonQ and Aer (CPU) across all qubit counts, with significantly lower execution times. For \num{28} qubits, Aer (GPU) completes in \SI{49}{\second}, achieving an \num{87}$\times$ speedup over IonQ and a \num{10}$\times$ speedup over Aer (CPU) at \num{28} qubits, highlighting the advantages of GPU acceleration.

To demonstrate Pilot-Quantum's capability to orchestrate parallel execution across physical quantum hardware, we project the performance of \num{256} QPUs processing \num{1024} random circuits. Due to the high cost and limited availability of quantum hardware, we executed only \num{8} random circuits per qubit configuration on IBM Eagle. The projected end-to-end time is estimated as the measured mean runtime scaled by a factor of \num{4} ($\num{1024}/\num{256}=\num{4}$), under the following assumptions: (i)~perfect linear scaling across \num{256} concurrently active QPUs with exclusive access (no queuing or contention), (ii)~independent, homogeneous task execution, and (iii)~the \num{8}-circuit sample is representative of the full workload distribution. These assumptions represent an idealized lower bound; actual performance would include queue wait times and backend variability. The time additionally accounts for the Pilot-Quantum scheduling overhead of \SI{2}{\second} for \num{1024} tasks on a single node. In contrast to simulation, the projected execution time does not grow exponentially with qubit count, as circuit execution time on a QPU is determined by circuit depth and gate count rather than state vector size.

\subsubsection{Distributed State Vector Simulations}\label{sec:dist_state_vector} 

\begin{figure}[t]
  \centering
  \resizebox{0.8\textwidth}{!}{
  \includegraphics{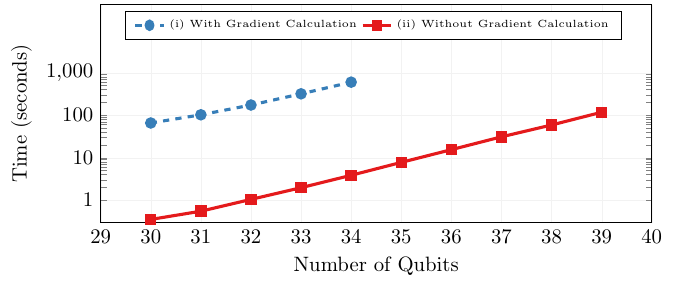}}
  \up
  \caption{\textbf{PennyLane \texttt{lightning.gpu} Distributed State Vector Simulations:} Computing the expectation value of a 2-layer strongly entangling layered (SEL) circuit with and without gradient calculation using PennyLane's \texttt{lightning.gpu} device. \up} 
  \label{fig:pq_mpi_dist_state_vector}
\end{figure}

Distributed state vector simulation enables efficient simulation of large quantum circuits with higher qubit counts and depth using multiple nodes and GPUs~\cite{DERAEDT201947, pennylane_dist_mem}. Figure~\ref{fig:pq_mpi_dist_state_vector} shows the distributed state vector simulation of a 2-layer strongly entangling layered (SEL) circuit~\cite{Schuld_2020}, which is frequently utilized for classification tasks. We compare circuit execution with and without gradient computation. We employ adjoint differentiation~\cite{jones2020efficientcalculationgradientsclassical} for gradient calculation, a method optimized for quantum simulations that offers lower memory and computational overhead than alternatives such as finite differences, which require more circuit evaluations. We investigate qubit counts from \num{30} to \num{34} qubits. The overhead of gradient calculation limits us to \num{34} qubits, whereas without gradient computation, we can scale up to \num{39} qubits. The results indicate that classical resources must be allocated appropriately between simulation and other computationally intensive tasks, such as ML, to ensure optimal performance.

\subsection{QML Workflow}
\label{sec:qml}

Quantum Machine Learning (QML) is a key workload~\cite{10.1145/3655027} enabling error mitigation and new applications~\cite{saurabh2023conceptual}. QML workflows typically include data preparation, training, and evaluation, leveraging parallelism for tasks such as hyperparameter optimization and architecture search. Data parallelism processes training subsets simultaneously, while tensor parallelism ensures efficient computation of gradients and backpropagation. Future workflows will likely integrate QML models for inference with other tasks. This section presents a QML mini-app with a multi-stage pipeline for data encoding, compression, and classifier training using Pilot-Quantum.

\begin{figure}[t]
  \centering
   \resizebox{0.8\textwidth}{!}{
   \includegraphics{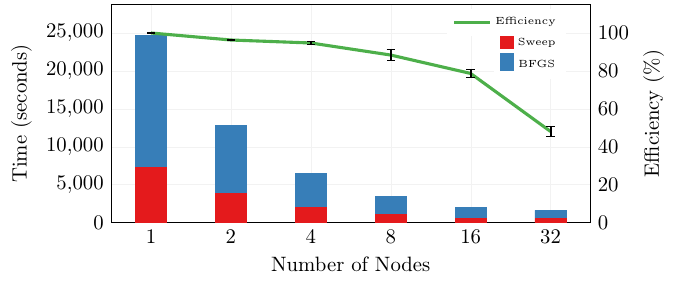}}\upp \upp
  \caption{\textbf{Execution time and efficiency} for compressing the 60,000 samples of the CIFAR-10 data set for cluster configurations with varying numbers of nodes. The time is divided between the two phases, with sweeping represented by the red portion and BFGS by the blue. 
  The results demonstrate that as the number of nodes increases, the overall execution time decreases while the efficiency reduces slightly, particularly for larger node counts. 
  \up}
  \label{fig:benchmark_qml}
\end{figure}

\emph{Data Encoding and Compression:}
\label{sec:compression}
Efficiently representing classical data (e.\,g., images) for QML is critical due to quantum state preparation bottlenecks~\cite{Aaronson:2015scy, Schuld2021}. Tensor networks, like matrix product states (MPS), can efficiently approximate images~\cite{jobst2023efficientmpsrepresentationsquantum}. Our QML pipeline mini-app uses Pilot-Quantum (Ray) for a two-stage data compression process.
We evaluated the pipeline using CIFAR-10~\cite{krizhevsky2009learning} comprising \num{10} classes and \num{60000} images, encoding each \(32 \times 32\) pixel image with \num{3} RGB color channels into a \num{13}-qubit quantum state (\num{10} qubits encoding spatial information, and \num{3} encoding color channels~\cite{rgba_encoding}). The image's quantum state is approximated by a depth \num{4} circuit with \num{205} trainable parameters, using only single and two-qubit gates on neighboring circuits. The first stage trains the circuit using a sweeping algorithm~\cite{Rudolph_2024}, while the second stage employs parameterized circuit training with a state vector simulator and BFGS optimizer~\cite{nocedal_numerical}.

\begin{figure}[t]
  \centering
  \resizebox{0.8\textwidth}{!}{
  \includegraphics{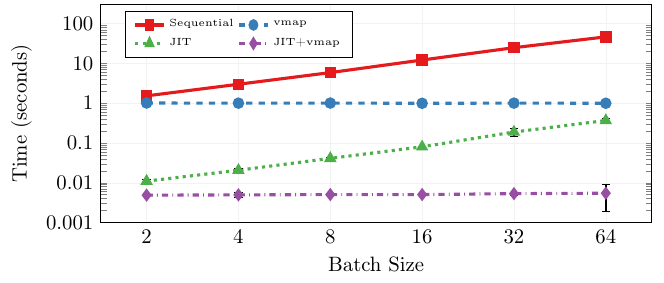}}\upp \upp
  \caption{\textbf{Batch processing time} by batch size for a variational quantum classifier implemented in PennyLane with JAX. The figure compares four processing methods: sequential (no optimizations), batch optimization using vmap, JIT compilation, and a combination of both. JIT compilation improves runtime by about two orders of magnitude. Batch optimization using vmap decouples the execution time from batch size. The combination of both vmap and JIT provides the best performance, significantly reducing processing time across all batch sizes. 
  }\upp \upp \upp
  \label{fig:training_loop_time}
\end{figure}

To demonstrate Pilot-Quantum's capabilities, we benchmarked the CIFAR-10 dataset compression runtime and efficiency across various cluster configurations (Figure~\ref{fig:benchmark_qml}). We used one ray worker node per physical node, each with \num{256} logical CPUs, and observed that throughput peaked at \SI{4.5}{tasks\per\second} with \num{256} CPUs. It indicates performance improvement, with up to \num{15}$\times$ speedup for the \num{32}-node cluster. However, efficiency decreased as the number of nodes increased. Efficiency, measured by task parallelization scaling with node count, is calculated as the N-node cluster runtime normalized to the single-node runtime and node count. The efficiency decrease stems from the sequential registration of ray worker nodes, causing initial processes to finish before the final node registers.

\emph{Classification:}
The compressed dataset was classified using a PennyLane-implemented variational quantum classifier. The circuit uses \num{13} qubits, with one arbitrary rotation layer per qubit and a ring of CNOTs~\cite{MariaSchuld2019}. Next, \num{10} Pauli-Z expectation values are used to generate the prediction. The ADAM optimizer trains the model by minimizing the cross-entropy loss between the predicted output and the target labels.

Figure~\ref{fig:training_loop_time} compares batch processing times for the variational quantum classifier using four methods: sequential (unoptimized), vmap batch optimization, JIT compilation, and combined vmap-JIT. JIT compilation improves runtime by about two orders of magnitude, while vmap decouples execution time from batch size. The vmap-JIT combination performs best, significantly reducing processing time across all batch sizes. This QML mini-app demonstrates how parallelism through Pilot-Quantum (Ray) and optimization tools, such as JAX, can significantly enhance QML workflow efficiency and scalability. 

\subsection{Q-Dreamer: Circuit Cutting Resource Optimizer}
\label{sec:circuit_cutting}

This section evaluates the Circuit Cutting Resource Optimizer introduced in section~\ref{sec:cc_resource_optimizer}. We use the \emph{EfficientSU2} ansatz, a variational circuit comprising single-qubit rotations and CNOT gates between adjacent qubits. Qiskit's expectation value reconstruction combines results from subcircuit executions.

\begin{figure}[t]
  \centering
  \resizebox{1\textwidth}{!}{
      \includegraphics{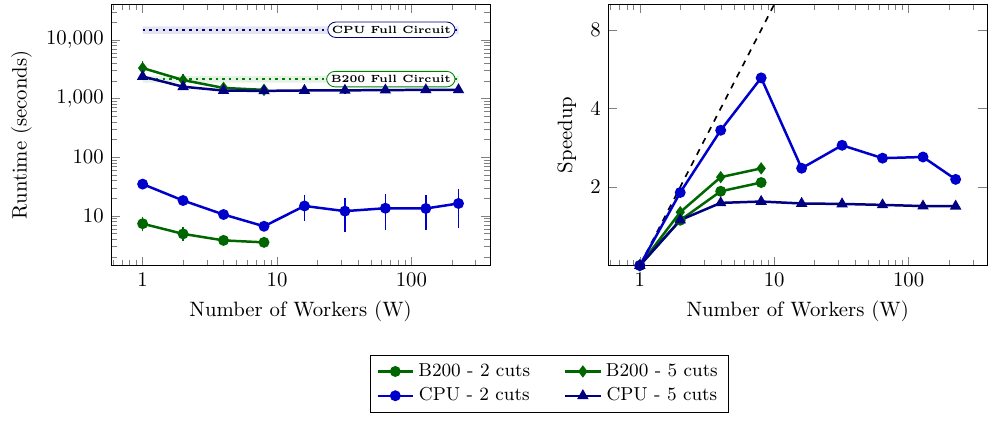}
  }
  \caption{\textbf{Circuit Cutting Strong Scaling for 36 Qubit EfficientSU2 Circuit:} Performance evaluation showing runtime and speedup characteristics as a function of the number of workers for circuit cutting operations.}
  \label{fig:qdreamer_worker_scaling}
\end{figure}

\emph{Circuit Cutting Runtime and Speedup:}
We evaluate runtime and speedup for a \num{36}-qubit EfficientSU2 circuit with two cutting scenarios: \num{2} cuts ($\approx \num{100}$ subexperiments) and \num{5} cuts ($\approx \num{46000}$ subexperiments). We use full circuit simulation as baseline.

Figure~\ref{fig:qdreamer_worker_scaling} shows that optimal performance varies dramatically across configurations: the \num{2}-cut scenario achieves near-ideal speedup with \num{2} GPU workers, while the \num{5}-cut scenario requires \num{8} workers; CPU execution shows diminishing returns beyond \num{56} workers. This complex optimization landscape motivates the need for predictive workload models.

\emph{Circuit Cutting Resource Optimizer (CCRO) Evaluation:}
We evaluate the CCRO using \num{165} measurements collected across NVIDIA B200/H100 GPUs (\num{80} data points) and multi-core CPUs (\num{85} data points), varying the number of cuts from \num{1} to \num{5} and qubit sizes (\num{34} and \num{36} qubits). We employ stratified train/test splits at ratios of \num{20}\%, \num{50}\%, and \num{70}\%. The efficiency parameters ($\eta_{\max}$, $p$) are fitted as described in section~\ref{sec:cc_resource_optimizer}.

Table~\ref{tab:optimal-cut-accuracy} reports the optimal cut selection accuracy, measuring how often the model correctly identifies the number of cuts that maximizes speedup. With 70\% training data, the general model achieves an average accuracy of 82\%. CPU predictions achieve high accuracy (88--93\%) across all configurations, while GPU predictions improve substantially with more training data, rising from 25\% to 75\%. Device-specific models perform comparably, suggesting that the general model captures cross-platform behavior effectively.

\begin{table}[t]
  \centering
  \caption{\textbf{Optimal Cut Selection Accuracy:} Percentage of test cases where the model correctly identifies the number of cuts that maximizes speedup. The general model is trained on combined GPU and CPU data and evaluated on all, GPU-only, or CPU-only test sets. Device-specific models are trained and tested exclusively on their respective hardware.}
  \label{tab:optimal-cut-accuracy}
  \small
  \begin{tabular}{@{}l *{3}{wc{1.5cm}} @{\hspace{1.2em}} *{2}{wc{1.5cm}} @{}}
  \toprule
  & \multicolumn{3}{c}{\textbf{General Model}} & \multicolumn{2}{c}{\textbf{Device-Specific Model}} \\
  \cmidrule(lr){2-4} \cmidrule(lr){5-6}
  \textbf{Train} & All & GPU & CPU & GPU & CPU \\
  \midrule
  20\% & 68.0\% & 25.0\% & 88.2\% & 25.0\% & 88.2\% \\
  50\% & 77.3\% & 50.0\% & 92.9\% & 50.0\% & 92.9\% \\
  70\% & \textbf{82.4\%} & 75.0\% & 88.9\% & 87.5\% & 77.8\% \\
  \bottomrule
  \multicolumn{6}{@{}l@{}}{\footnotesize Data: 165 points (80 GPU [B200, H100], 85 CPU), Circuits with 34 and 36 qubits.} \\
  \end{tabular}
\end{table}

\begin{figure}[t]
  \centering
  \resizebox{\textwidth}{!}{
      \includegraphics{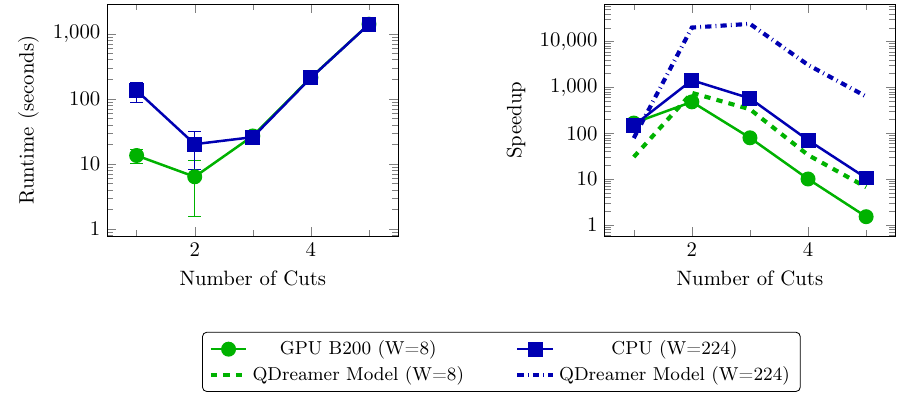}
  }
  \caption{\textbf{Circuit Cutting Performance and Q-Dreamer Estimations for a 36-qubit EfficientSU2 Circuit:} (a)~Runtime versus number of cuts for GPU (B200, $W{=}8$) and CPU ($W{=}224$) backends. (b)~Measured speedup (markers) compared to Q-Dreamer model predictions (lines). Both configurations achieve peak speedup at $k{=}2$ cuts, with Q-Dreamer correctly identifying the optimal operating point. The model captures the characteristic tradeoff between subcircuit size reduction ($2^{n-n_{\text{sub}}}$) and sampling overhead ($9^k$).}
  \label{fig:qdreamer_cc_performance}
\end{figure}

Figure~\ref{fig:qdreamer_cc_performance} evaluates the 50\% model (calibrated using approximately $\approx$82 measurements) on GPU ($W{=}8$ B200) and CPU ($W{=}224$ cores) backends. The left panel shows measured runtime versus the number of cuts; the right panel compares measured speedup (markers) with model predictions (dashed lines). The results demonstrate that the optimizer accurately captures speedup behavior: performance initially improves with more cuts due to reduced subcircuit complexity, peaks at $k{=}2$ for both configurations, and then degrades as the $9^k$ sampling overhead dominates.

\emph{Discussion and Limitations:}

Across the evaluation, the model correctly identifies the cut count that maximizes speedup in 70--93\% of test cases depending on train split and device (table~\ref{tab:optimal-cut-accuracy}). Nevertheless, the model can overestimate speedup due to unmodeled per-task overheads that compound with $9^k$ (e.\,g., reconstruction, task dispatch), tail latency from stragglers, and GPU kernel launch overheads. Limitations include: (i)~the power-law efficiency model breaks down for highly imbalanced cut workloads; (ii)~parameters calibrated on EfficientSU2 circuits may not transfer to other circuit families; (iii)~distributed multi-node scaling introduces communication overhead not captured in the current formulation; and (iv)~the model optimizes for performance without considering fidelity tradeoffs, where increased cuts amplify sampling error. Future work should incorporate straggler-aware cost modeling, fidelity constraints, and evaluation across broader circuit types and scales (including QPUs).

\section{Conclusion and Future Work} 
\label{sec:discussion}

This paper addresses a central challenge in quantum-HPC integration: bridging the gap between quantum applications and their efficient execution on hybrid quantum-classical infrastructure. Our four contributions encompass conceptual architecture, characterization methodology, middleware implementation, and workload optimization, enabling researchers and practitioners to develop, deploy, and optimize hybrid quantum-classical applications while abstracting the underlying complexities of resource management.

The practical impact is threefold. First, the execution motifs and mini-apps provide a common vocabulary for reasoning about hybrid application requirements, enabling systematic middleware design rather than ad-hoc solutions. Second, Pilot-Quantum demonstrates that the pilot abstraction, proven effective in classical distributed computing, extends naturally to quantum-HPC environments, achieving significant speedups. Third, Q-Dreamer shows that analytical performance models can accurately predict optimal resource configurations for circuit-cutting workloads, correctly identifying the tradeoff between subcircuit parallelism and reconstruction overhead.

\emph{Limitations:} Pilot-Quantum and Q-Dreamer are proof-of-concept implementations designed to understand the resource management challenges arising in quantum-HPC systems. Pilot-Quantum currently has three limitations: (i) task dependencies are handled only at the application level; (ii) it lacks high-level workflow representations such as DAGs; and (iii) QPU access is mediated by high-level libraries (e.\,g., PennyLane, Qiskit), which limits low-level control.

As quantum hardware scales and new algorithmic patterns emerge, the modular architecture enables targeted extensions without requiring redesign of the entire stack. The analytical models in Q-Dreamer can also serve as an evaluator environment for reinforcement learning and other ML-based approaches to workload optimization, enabling data-driven improvements to scheduling policies.

\emph{Future Work:}
We envision three future work directions: First, extending Q-Dreamer to additional execution motifs (e.\,g., variational algorithms, error mitigation, distributed simulation) and integrating physical QPU characteristics into fidelity-aware scheduling that considers, e.\,g., gate error rates. Second, developing cross-layer optimization strategies in which workflow-level algorithm selection informs resource-level scheduling and resource performance characteristics influence workload parallelization. Third, leveraging the mini-app methodology to establish standardized benchmarks for quantum-HPC middleware, enabling objective performance comparisons across implementations.

\subsubsection*{Acknowledgements}
This research used resources from the Oak Ridge Leadership Computing Facility, a DOE Office of Science User Facility supported under Contract DE-AC05-00OR22725 (Project CSC595).  Additionally, the work received support from NERSC (ERCAP0032477/ERCAP0029512: "Characterization and Middleware for Hybrid Quantum-HPC Applications") and the Bavarian State Ministry of Economic Affairs (BenchQC project, Grant DIK-0425/03). The authors used OpenAI's, Anthropic's, and Google's language-generation models to generate parts of the text, which were subsequently reviewed, edited, and revised. 

\printbibliography 

\end{document}